\documentclass[english,aps,prb,amsmath,amssymb,reprint]{revtex4-1}
\usepackage[T1]{fontenc}
\usepackage[latin9]{inputenc}
\usepackage{xcolor}
\usepackage{babel}
\usepackage{textcomp}
\usepackage{amstext}
\usepackage{graphicx}
\PassOptionsToPackage{normalem}{ulem}
\usepackage{ulem}
\usepackage[unicode=true,
 bookmarks=true,bookmarksnumbered=false,bookmarksopen=false,
 breaklinks=false,pdfborder={0 0 0},backref=false,colorlinks=true,pdfpagemode=FullScreen]
 {hyperref}
\hypersetup{pdftitle={Pentagons in the Si(331)-(12x1) surface reconstruction},
 pdfauthor={Ruslan Zhachuk and Sergey Teys},
 pdfkeywords={Silicon, Surface, Structure, STM, DFT},
 allcolors=blue}
\usepackage{breakurl}

\makeatletter

\newcommand{\noun}[1]{\textsc{#1}}
\providecolor{lyxadded}{rgb}{0,0,1}
\providecolor{lyxdeleted}{rgb}{1,0,0}

 \@ifundefined{textcolor}{}
 {%
   \definecolor{BLACK}{gray}{0}
   \definecolor{WHITE}{gray}{1}
   \definecolor{RED}{rgb}{1,0,0}
   \definecolor{GREEN}{rgb}{0,1,0}
   \definecolor{BLUE}{rgb}{0,0,1}
   \definecolor{CYAN}{cmyk}{1,0,0,0}
   \definecolor{MAGENTA}{cmyk}{0,1,0,0}
   \definecolor{YELLOW}{cmyk}{0,0,1,0}
 }

\makeatother

\begin{document}

\title{Pentagons in the Si$(331)-(12\times1)$ surface reconstruction}

\author{Ruslan Zhachuk}

\email{zhachuk@gmail.com}

\affiliation{Institute of Semiconductor Physics, pr. Lavrentyeva 13, Novosibirsk
630090, Russia}

\author{Sergey Teys}

\affiliation{Institute of Semiconductor Physics, pr. Lavrentyeva 13, Novosibirsk
630090, Russia}

\author{\today}

\begin{abstract}
The microscopic structure of the high-index Si(331) \textminus{} (12
\texttimes{} 1) surface is investigated combining scanning tunneling
microscopy with \emph{ab initio} calculations. We present a new structural
model of the Si(331) surface, employing a novel reconstruction element
composed of six pentagons integrated to the structure of the adjacent
pentamer with an interstitial atom. We demonstrate that appropriately
arranged additional pentagons significantly lower the surface energy
of the high-index surface. The model predicts the existence of multiple
Si(331) buckled configurations with similar energies.
\end{abstract}
\maketitle
High-index surfaces of Si are interesting for both fundamental research
and technological applications. The technological interest is based
on the demonstrated improved heteroepitaxial growth on such surfaces
and the use of them as templates for nanostructures growth.\cite{zha04}
Such surfaces, however, often demonstrate complex surface reconstructions.
The problem of finding the atomic structure of surface reconstructions
is still a formidable challenge. The main difficulty is the existence
of a large number of atomic configurations for surface cells even
with a moderate number of atoms. Scanning tunneling microscopy (STM)
and density functional theory (DFT) calculations are two complementary
methods often used in conjunction for surface structure determination.
Although DFT calculations offer accurate total energies, the surface
structure prediction of materials with large surface cells is very
hard nowadays due to a high computational cost of such calculations.
The experimental STM data help a lot to narrow down the search for
possible atomic configurations by showing the actual structure of
a surface at the atomic scale of a real sample. However, the interpreting
of high resolution STM images can be very tricky, since STM does not
actually show positions of atomic nuclei. In the most simplified view,
the STM images represent a mixture of surface topography and a map
of local density of electronic states of a sample surface.\cite{hof03,ter85}
Consequently, the interpreting of such images, in its part, may require
the knowledge of surface atomic structure and \emph{ab initio} calculations.

\begin{figure}
\includegraphics[clip,width=6.5cm]{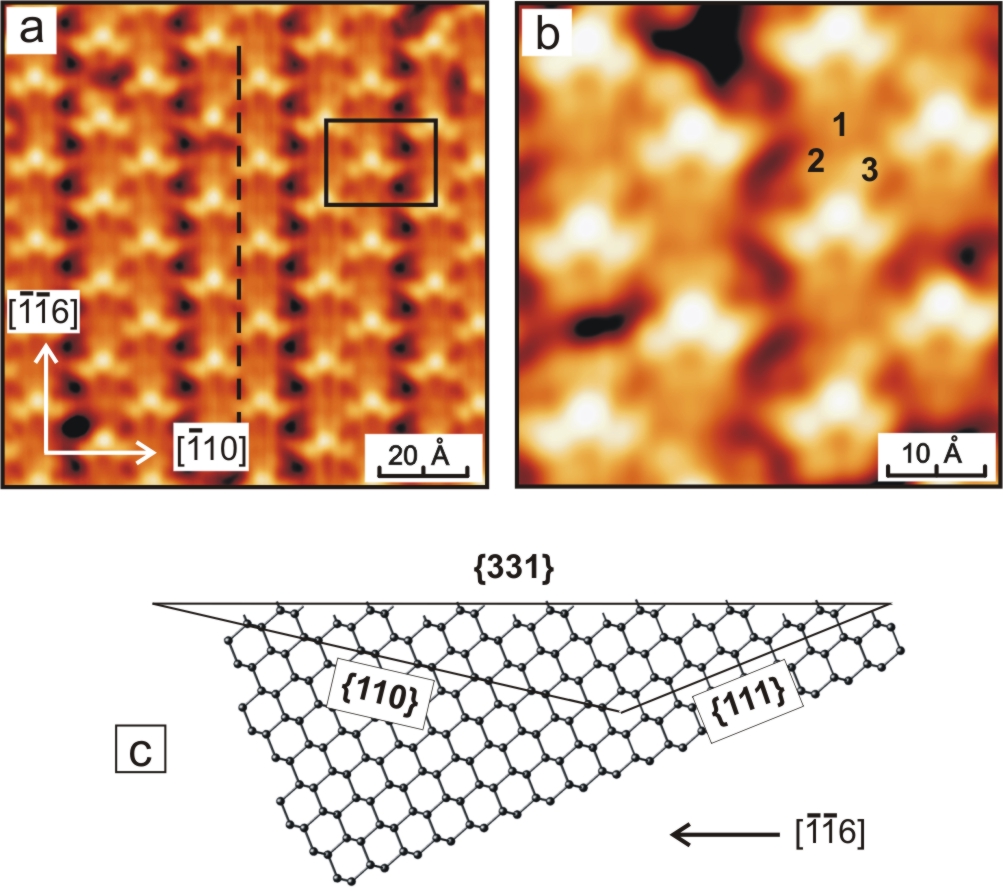}

\caption{\label{fig1}(color online). (a), (b) Experimental STM images of the
Si$(331)-(12\times1)$ surface. (a) $U=+1.0\,\mathrm{V}$, $I=0.03\,\mathrm{nA}$.
The calculated unit cell is outlined. The orientation of the glide
plane is indicated by a dashed line. (b) $U=+0.8\,\mathrm{V}$, $I=0.024\,\mathrm{nA}$.
The atoms, resolved between pentamers, are numbered $1-3$. (c) A
side view of the silicon crystal lattice in the $(\bar{1}10)$ plane.
$(111)$, $(110)$ and $(331)$ planes are marked. }
\end{figure}

Si$(331)$ is a flat silicon surface exhibiting a complex reconstruction.
The surface structure is often designated as $(12\times1)$ or $(6\times2)$,
although the correct notation can only be given by matrix.\cite{bat09,olsh98}
The study of $(12\times1)$ surface reconstruction has long history.
Three structural models were proposed.\cite{bat09,olsh98,gai01} It
was recognized from the very beginning that the rectangular surface
unit cell contains two identical structural units (Fig.~\ref{fig1}(a)).\cite{hib93}
The first structural unit is located at the surface cell corner. The
second unit is shifted by $a/2$ from the center to $[\bar{1}10]$
or $[1\bar{1}0]$, where $a$ is a basic translational unit of the
unreconstructed $(331)$ plane in that direction. The surface has
a glide plane symmetry along the $[\bar{1}\bar{1}6]$ direction running
through the center of the zigzag chain of structural units (dashed
line in Fig.~\ref{fig1}(a)). 

There were several attempts to construct the observed structural units
from the elementary building blocks known from the previous studies
of silicon surfaces.\cite{olsh98,gai01,bat09} It was proposed that
the structural units consist of adatoms\cite{olsh98} or adatoms and
dimers.\cite{gai01} In the most recent structural model proposed
by Battaglia \emph{et al.},\cite{bat09} those units were constructed
from the pentamer with an interstitial atom (hereafter pentamer) and
two adatoms. Originally, the pentamers were suggested as a structural
building block on the silicon $(113)$ surface\cite{dab95} and were
used to explain the structure of Si$(110)$ later.\cite{ste04} The
model by Battaglia \emph{et al.}\cite{bat09} basically represents
an adaptation of the adatom-tetramer-interstitial (ATI) model of the
Si$(110)-(16\times2)$ surface reconstruction by Stekolnikov \emph{et
al.}\cite{ste04} for the Si$(331)-(12\times1)$ surface. We, therefore,
refer to the structural model proposed in Ref.~\onlinecite{bat09}
as the ATI model. It was demonstrated that the pentamers indeed adequately
describe the groups of five bright spots observed in the experimental
STM images of the Si$(331)$ surface.\cite{bat09} Nevertheless, the
ATI model of Si$(331)-(12\times1)$ is questionable as it shows a
poor agreement with STM images of the areas between the pentamers
and it leads to the high surface energy, as demonstrated below.

The aim of our work is to develop a relistic Si$(331)-(12\times1)$
surface reconstruction model by a combined experimental and theoretical
study. We propose a microscopic model of the $(12\times1)$ reconstruction
which shows a remarkably low surface energy and explains the experimental
STM data. 

The STM images were recorded at room temperature in the constant-current
mode using an electrochemically etched tungsten tip. The measurements
were performed in an ultrahigh vacuum chamber ($7\times10^{-11}\,\mathrm{Torr}$)
on a system equipped with an Omicron STM. A clean Si$(331)$ surface
was prepared by sample flash annealing at $1250\,\mathrm{{^\circ}C}$
for one minute followed by stepwise cooling with $2\,\mathrm{{^\circ}C}$
per minute steps within temperature range $400\text{\textminus850}\,\mathrm{{^\circ}C}$.
More details on the experimental procedure can be found in Ref.~\onlinecite{zha13}.
The \textsc{WSXM} software was used to process the experimental and
calculated STM images.\cite{hor07}

The calculations were carried out using the pseudopotential\cite{tro91}
DFT \textsc{siesta} code\cite{sol02} within the local density approximation
to the exchange and correlation interactions between electrons.\cite{per92}
The valence states were expressed as linear combinations of numerical
atomic orbitals of the Sankey-Niklewski type.\cite{sol02} In the
present calculations, the polarized double-$\mathrm{\zeta}$ functions
were assigned for all species. This means two sets of\emph{\noun{
}}$s$ and $p$ orbitals plus one set of $d$ orbitals on Si atoms,
and two sets of $s$ orbitals plus a set of $p$ orbitals on H. The
electron density and potential terms were calculated on a real space
grid with the spacing equivalent to a plane-wave cut-off of $200\,\mathrm{Ry}$.

The surface energy (per unit area) of the reconstructed Si$(331)$
surface ($\gamma_{rec}$) was calculated as $\gamma_{rec}=\gamma_{unrec}+\triangle\gamma_{rec}$,
following the procedure described in Refs.~\onlinecite{ste02, zha13}.
Here $\gamma_{unrec}$ is the energy of the unreconstructed and unrelaxed
Si$(331)$ surface, and $\triangle\gamma_{rec}$ is the energy gain
due to surface reconstruction and relaxation. $\mathrm{\gamma_{unrec}}$
was calculated using a symmetric slab, 20 Si bilayers thick. $\mathrm{\triangle\gamma_{rec}}$
were calculated using 10 bilayers thick slabs terminated by hydrogen
from one side. A $10\,\mathrm{\mathring{A}}$ thick vacuum layer was
used. The rectangular surface unit cell, as outlined in Fig.~\ref{fig1}(a),
was employed. The Brillouin zone was sampled using a $4\times4\times1$
$\mathbf{k}$-point grid.\cite{mon76} The geometry was optimized
until all atomic forces became less than $1\,\mathrm{meV/\mathring{A}}$.
The constant-current STM images were produced within the Tersoff-Hamann
approach\cite{ter85} using eigenvalues and eigenfunctions of the
Kohn-Sham equation\cite{koh65} for a relaxed atomic structure.

The tests were carried out to monitor the convergence of simulated
STM images and surface energies with respect to basis set, Brillouin
zone integration, slab thickness and separation between slabs. We
estimate an error less than $1\,\mathrm{meV/\mathring{A}^{2}}$ for
the calculated surface energy differences between relaxed structures.
The absolute values of surface energies are overestimated by about
$3\text{\textminus7}\,\mathrm{\mathrm{meV/\mathring{A}^{2}}}$.

The ATI structural model by Battaglia \emph{et al.}\cite{bat09} has
two main flaws. First, the calculated surface energy, according to
that model, is too high. The upper limit for the Si$(331)$ surface
energy can be estimated by requiring surface stability to faceting
to Si$(111)$ and Si$(110)$. All three planes are schematically shown
in Fig.~\ref{fig1}(c). Therefore,
\begin{equation}
\mathrm{\Gamma_{(331)}S_{(331)}=\gamma_{(111)}S_{(111)}+\gamma_{(110)}S_{(110)}},\label{eq:1}
\end{equation}
where $\mathrm{\Gamma_{(331)}}$ is the upper limit for the Si$(331)$
surface energy, $\mathrm{\gamma_{(111)}}$ and $\mathrm{\gamma_{(110)}}$
are surface energies for Si$(111)$ and Si$(110)$, respectively.
$\mathrm{S_{(331)}}$, $\mathrm{S_{(111)}}$, $\mathrm{S_{(110)}}$
are the surface areas of $(331)$, $(111)$ and $(110)$, which are
mutually dependent due to geometrical constraints (Fig.~\ref{fig1}(c)):
$\mathrm{S_{(110)}\approx0.649\cdot S_{(331)}}$, $\mathrm{S_{(111)}\approx0.397\cdot S_{(331)}}$.
The surface energy of Si$(111)-(7\times7)$, according to the dimer-adatom
stacking fault model by Takayanagi \emph{et al.},\cite{tak85} is
$84.9\,\mathrm{meV/\mathring{A}^{2}}$,\cite{ste02} while the surface
energy of Si$(110)-(16\times2)$ is $103.7\,\mathrm{meV/\mathring{A}^{2}}$
according to the structural model by Stekolnikov \emph{et al.}\cite{ste04}
Thus, the estimated upper limit for the Si$(331)-(12\times1)$ surface
energy according to the Eq.~\ref{eq:1} is $101.0\,\mathrm{meV/\mathring{A}^{2}}$,
which is $\approx7\,\mathrm{meV/\mathring{A}^{2}}$ less than the
value given in Ref.~\onlinecite{bat09a}. This means that, according
to the ATI model of the Si$(331)$ surface, it should be decomposed
into Si$(111)$ and Si$(110)$ facet surfaces in the obvious contradiction
with experiments. 

\begin{figure}
\includegraphics[clip,width=6.5cm]{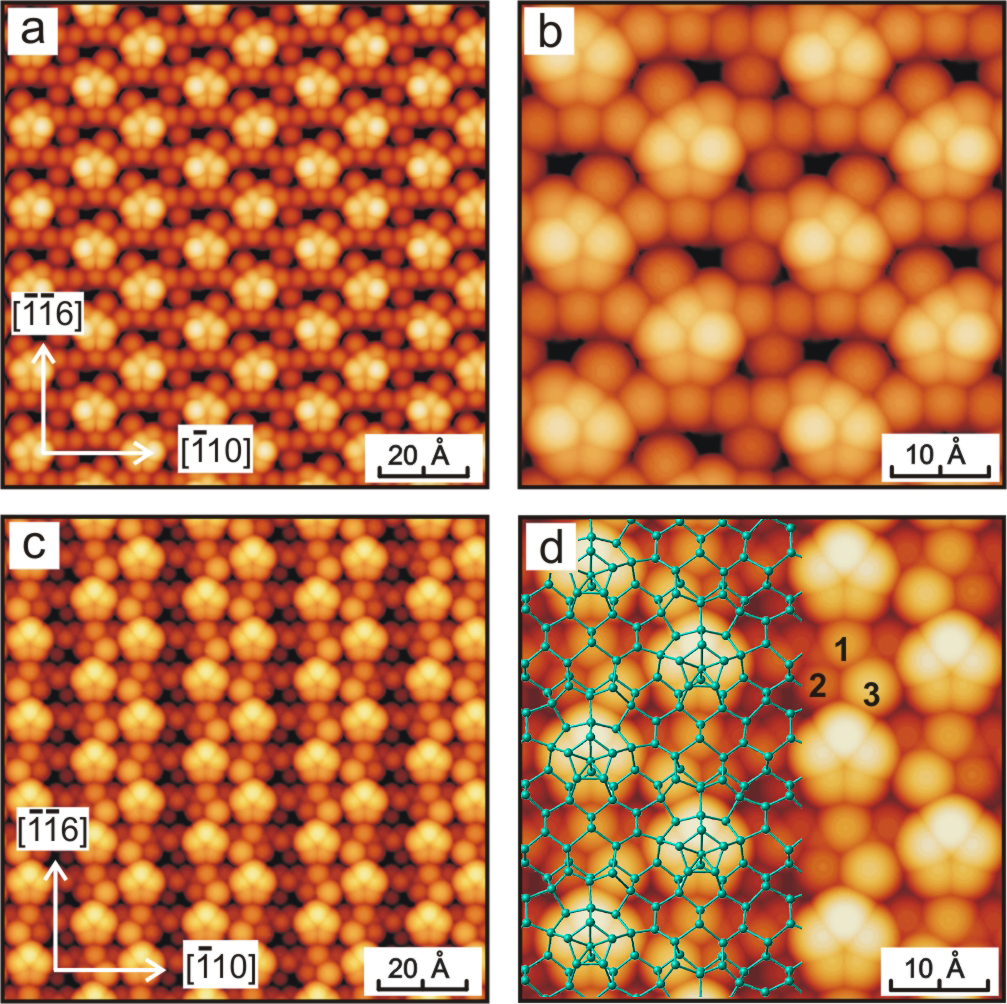}

\caption{\label{fig2}(color online). (a), (b) Calculated STM images of the
Si$(331)-(12\times1)$ surface assuming the ATI atomic model by Battaglia
\emph{et al.} \cite{bat09} (c), (d) Calculated STM images of the
Si$(331)$ surface assuming the 8P atomic model for the $(12\times1)$
reconstruction proposed in the present study. The atoms between pentamers,
resolved in the experimental STM images, are numbered $1-3$ in (d).
The 8P atomic model of the $(12\times1)$ reconstruction is superimposed
in the STM image in (d). Bias voltage corresponds to $+0.8\,\mathrm{eV}$
with respect to the theoretical Fermi level for all calculated STM
images (empty electronic states). See Supplementary Fig. 3 for a filled
states calculated STM image of the Si$(331)-(12\times1)$.}
\end{figure}

Second, our \emph{ab initio} investigation demonstrates that the relaxed
ATI model by Battaglia \emph{et al.} cannot account for the important
surface features observed in experiments. The calculated constant-current
STM images of Si$(331)-(12\times1)$, based on the ATI structural
model, are shown in Figs.~\ref{fig2}(a) and \ref{fig2}(b). The
pentamers, indeed, reproduce the brightest STM image features in Figs.~\ref{fig1}(a)
and \ref{fig1}(b). On the other hand the vertical dark stripes in
the $[\bar{1}\bar{1}6]$ direction clearly visible in experimental
STM images, are not reproduced. The dark stripes, representing surface
depressions or trenches, have been observed almost in every STM study
of the Si$(331)$ surface and, therefore, the correct structural model
should account for this surface feature.\cite{bat09,gai01,hib93}
All these problems - incorrect STM images and too high surface energy
- taken together imply that the ATI model of the $(12\times1)$ by
Battaglia \emph{et al.} is not a good model for Si$(331)$.

\begin{figure}
\includegraphics[clip,width=6.5cm]{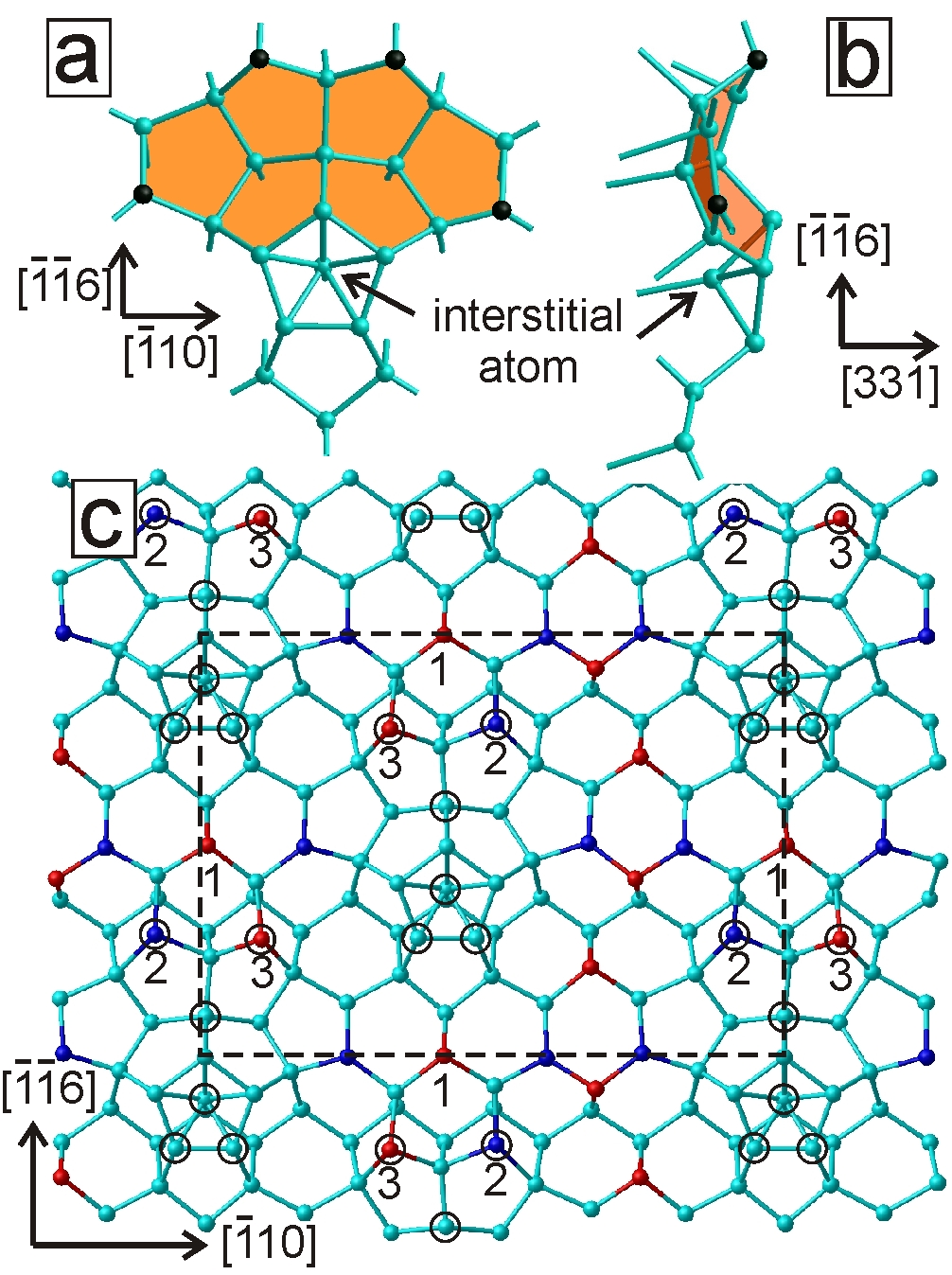}

\caption{\label{fig3}(color online). (a), (b) The elementary building block
structure of the Si$(331)-(12\times1)$ surface: 8-pentagons unit
(8PU). Only saturated bonds are shown. The atoms with dangling bonds
are marked in black. The pentagons in 6PU are highlighted in orange
for eye guidance purposes. (a) Plan view. (b) Side view. (c) The 8P
model for the Si$(331)-(12\times1)$ surface reconstruction. The atomic
positions after surface relaxation are shown. The unit cell is outlined
by a dashed line. Red/blue balls indicate raised/lowered under-coordinated
Si atoms. Black circles indicate the additional atoms in relation
to the unreconstructed Si$(331)$ surface. The atoms between pentamers,
resolved in STM, are numbered $1-3$.}
\end{figure}

The new structural building block, proposed in this work, is shown
in Figs.~\ref{fig3}(a) and \ref{fig3}(b). It contains a 6-pentagon
unit (6PU) and the pentamer with an interstitial atom. The 6PU structure
can be represented as two mirror-symmetrical groups with three pentagons
in each of them (3-pentagon unit, 3PU). The pentagons in 3PU are folded
into a trefoil with one of its lobes being side of the pentamer structure.
This makes 6PU to be closely integrated into the structure of the
adjacent pentamer. The silicon interconnections in 3PU are similar
to that in $\mathrm{C{}_{20}}$ - the smallest fullerene.\cite{jar00}
The 3PU surface is concaved, like the $\mathrm{C{}_{20}}$ surface,
if viewed from the inside of a fullerene. The 6PU, as shown in Figs.~\ref{fig3}(a)
and \ref{fig3}(b), has only four dangling bonds (four under-coordinated
Si atoms). The pentamer with an interstitial atom introduces two additional
pentagons: one at the top of the pentamer and the other - on the side
away from 6PU. Therefore, we refer to the complete structure, composed
of a pentamer and 6PU, as an 8-pentagon unit (8PU). 

The atomic model of the Si$(331)-(12\times1)$ surface, composed of
8PUs and presented in Fig.~\ref{fig3}(c), is named 8P. The 8P model
has two less unsaturated bonds (under-coordinated Si atoms) per unit
cell, as compared to the ATI structural model proposed in Ref.~\onlinecite{bat09}.
According to the 8P structural model, only six additional Si atoms
per 8PU are required to form the $(12\times1)$ reconstruction on
the initially unreconstructed surface (these atoms are marked by black
circles in Fig.~\ref{fig3}(c)).

The ideal unrelaxed 8PU has a mirror symmetry in the $(\bar{1}10)$
plane (Fig.~\ref{fig3}(a)). This symmetric atomic configuration
is, however, unstable against buckling. When relaxing the structure,
the under-coordinated Si atoms are displaced either away (raised)
or toward the bulk (lowered), as marked by red/blue balls in Fig.~\ref{fig3}(c).
Similar structural transformations are well known for dimers on Si$(100)-(2\times1)$\cite{ram95}
and also have been observed for more complex structures on the triple
step edges of the Si$(7\,7\,10)$ surface.\cite{zha14,tey06} Thus,
the mirror symmetry of relaxed 8PU breaks due to buckling of surface
atoms, although the glide plane symmetry of the $(12\times1)$ reconstruction
along the $[\bar{1}\bar{1}6]$ direction retains. 

The three bonds of raised atoms become strongly $p$-like, and a fully
occupied dangling bond state, mostly $s$-like, is formed. Conversely,
the lowered atoms become approximately $sp^{2}$-coordinated. They
produce high-energy $p$-like dangling bond states, whose electrons
are donated to the $s$-type radicals on raised atoms. The raised/lowered
silicon atoms interact with each other due to a charge transfer between
them and the locally induced tensile/compressive strain. 

Due to the buckling of the surface atoms in 8PU, multiple configurations
of the $(12\times1)$ reconstruction are possible. There are 8 symmetry
nonequivalent atoms with dangling bonds per $(12\times1)$ unit cell
(Fig.~\ref{fig3}(c)). In the absence of the interaction between
them, their buckling would be uncorrelated and we could expect $2^{8}=256$
configurations with the glide plane symmetry. We have found, however,
only 8 atomic configurations which are, at least, metastable out of
98 (most probable) relaxed structures with a glide plane symmetry.
These configurations are shown in Supplementary Fig. 1. The surface
energies of most of them cluster in the $2\,\mathrm{meV/\mathring{A}^{2}}$
energy window. The mixed configurations \emph{ij}, composed of symmetric
configurations \emph{i} and \emph{j} are also metastable (see Supplementary
Fig. 2, for an example of such structure). These configurations have
no glide plane symmetry. The Si$(331)$ surface, in principle, should
adopt the configuration with the lowest energy. However, the influence
of the STM tip (electric field, injected charge) cannot be excluded
since the calculated structures are quasi-degenerate. The Si$(331)-(12\times1)$
surface configuration, which demonstrates the best agreement with
the experimental STM images, is shown in Fig.~\ref{fig3}(c) and
discussed below. 

Local and reversible modification of the buckled Ge$(100)$ atomic
structure by STM tip has been reported.\cite{tak04} The results have
been discussed in the context of realizing a rewritable nanometer-scale
memory.\cite{cho96} The existance of multiple buckled configurations
of the Si$(331)$ surface with similar energies imply that these effects
can be observed on Si$(331)$ as well. This idea deserves furher research. 

The formation energy of the unreconstructed and unrelaxed Si$(331)$
surface is $129.7\,\mathrm{meV/\mathring{A}^{2}}$ according to our
calculation. The energy gain due to the $(12\times1)$ surface reconstruction
and relaxation, according to the ATI model proposed by Battaglia \emph{et
al.},\cite{bat09} is $15.8\,\mathrm{meV/\mathring{A}^{2}}$ (our
data). Thus, the surface energy according to that model is $113.9\,\mathrm{meV/\mathring{A}^{2}}$.
These values are in a reasonable agreement with the data reported
in Ref.~\onlinecite{bat09a}. The energy gain due to the surface
reconstruction, according to the 8P model, shown in Fig.~\ref{fig3}(c),
is $31.2\,\mathrm{meV/\mathring{A}^{2}}$. Therefore, according to
the 8P model, the Si$(331)-(12\times1)$ surface energy is $15.4\,\mathrm{meV/\mathring{A}^{2}}$
lower than in the ATI model proposed in Ref.~\onlinecite{bat09}.
Such huge energy difference is far beyond the possible error in computed
relative surface energies. The surface energy of Si(331)-(12\texttimes{}1),
according to the 8P model, is $98.5\,\mathrm{meV/\mathring{A}^{2}}$,
which is below its estimated upper limit, calculated using the Eq.~\ref{eq:1}.
Moreover, the calculated surface energy is close to that of the Si$(111)-(7\times7)$,
which is $92.1\,\mathrm{meV/\mathring{A}^{2}}$ according to our results
obtained using a similar calculation procedure.\cite{zha13} The Si(111)-(7\texttimes{}7)
surface is, in turn, known to be the most stable silicon surface with
the lowest energy.\cite{ste02,ste05}

There are several reasons for the low surface energy of Si(331)-(12\texttimes{}1)
in the 8P model. First, the number of dangling bonds in the 8P model
is less than in the ATI model. Second, the bond lengths in 8P are
nearly the bulk bond length and they are less stretched than in the
ATI model. Third, the bond angles are only slightly distorted with
respect to the tetrahedral structure. Fourth, the surface energy is
additionally decreased due to the buckling of surface atoms.\cite{bech03}

The structure of 6PU is difficult to visualize in STM because most
of its bonds are saturated and its surface is concaved. The same difficulty
exists for the dimers in the Si$(111)-(7\times7)$ reconstruction,
which, to our knowledge, have never been observed in STM. The high-resolution
STM image of the Si$(331)$ surface exhibiting the $(12\times1)$
reconstruction is presented in Fig.~\ref{fig1}(b). The image agrees
with the study of Battaglia \emph{et al.,}\cite{bat09} but it reveals
more details between pentamers (Fig.~\ref{fig1}(b)). There are a
few surface defects visible in the presented STM image, but the repeating
structural units are easily recognized. Besides the pentamer structure,
clearly resolved in Fig.~\ref{fig1}(b), three symmetry nonequivalent
atoms can be distinguished in the experimental STM image. These atoms
are numbered $1-3$ in the experimental STM image in Fig.~\ref{fig1}(b),
in the calculated STM image in Fig.~\ref{fig2}(d) and in the atomic
model in Fig.~\ref{fig3}(c). Atom 3 is also visible in the STM images
by Battaglia \emph{et al.}\cite{bat09} and it was attributed to the
adatom in the ATI atomic model. According to the 8P model, however,
atoms 2 and 3 correspond to the under-coordinated buckled Si atoms
in the 6PU structure (atom 2 is lowered, atom 3 is raised), while
atom 1 is a rest-atom of the Si$(331)$ surface. 

The 8P model correctly reproduces the trenches in the $[\bar{1}\bar{1}6]$
direction as one can see in the large scale calculated STM image in
Fig.~\ref{fig2}(c). The trench area is located between the zig-zag
rows of 8PUs. Due to the 3D structure of 8PUs, the atoms in the trench
appear relatively lower (darker) in STM images. One may suggest that
the trenches serve for the strain relaxation introduced by 8PUs similar
to the dislocations formed in the strained systems during growth. 

In summary, we have presented a novel model of the Si(331) surface.
The new model consistently describes the experimental STM data and
demonstrates the remarkably low surface formation energy. The model
predicts that many surface configurations are possible depending on
the buckling states of Si(331) reconstruction elements. This can potentially
be used for information storage and requires further research. 
\begin{acknowledgments}
Gratefully acknowledged are the fruitful discussions with A. Shklyaev
and J. Coutinho. We would like to thank the Novosibirsk State University
for providing the computational resources. This work was supported
by the Russian Foundation for Basic Research (Project No. 14-02-00181).
\end{acknowledgments}
\bibliographystyle{apsrev4-1}

\end{document}